\documentclass[10pt]{article}
\usepackage[dvips]{graphicx}
\begin{document}
\begin{center}
\textbf{{\large Born-Infeld Type Phantom Model in the
$\omega-\omega'$ Plane}}
 \vskip 0.35 in
\begin{minipage}{4.5 in}
\begin{center}
{\small Z. G. HUANG$^\dag$, X. H. LI and Q. Q. SUN \vskip 0.06 in
\textit{ Department~of~Mathematics~and~Physics,
\\~Huaihai~Institute~of~Technology,~222005,~Lianyungang,~China
\\
$^\dag$zghuang@hhit.edu.cn}}
\end{center}
\vskip 0.2 in

{\small In this paper, we investigate the dynamics of
Born-Infeld(B-I) phantom model in the $\omega-\omega'$ plane, which
is defined by the equation of state parameter for the dark energy
and its derivative with respect to $N$(the logarithm of the scale
factor $a$). We find the scalar field equation of motion in
$\omega-\omega'$ plane, and show mathematically the property of
attractor solutions which correspond to $\omega_\varphi\sim-1$,
$\Omega_\varphi=1$, which avoid the "Big rip" problem and meets the
current observations well. \vskip 0.2 in \textit{Keywords:} Dark
energy; Born-Infeld; Phantom; $\omega-\omega'$ plane; Attractor.
}
\end{minipage}
\end{center}
\vskip 0.2 in
\begin{flushleft}\textbf{1. Introduction}\end{flushleft}
Recent observations of high-redshift Type Ia Supernova[1] and the
Cosmic Microwave Background[2] have shown us such a fact: the
density of clustered matter including cold dark matters plus
baryons, $\Omega_{m0}\sim1/3$, and that the Universe is flat to high
precision, $\Omega_{total}=0.99\pm0.03$[3]. That is to say, we are
living in a flat universe which it undergoing a phase of accelerated
expansion, and there exists an unclumped form of energy density
pervading the Universe. This unknown energy density which is called
"dark energy" with negative pressure, contributes to two thirds of
the total energy density. Perhaps the simplest explanation for these
data is that the dark energy corresponds to a positive cosmological
constant. However, the cosmological constant model suffers from two
serious issues called "coincidence problem" and "fine-tuning
problem". An alternative is a scalar field which has not yet reached
its ground state. These scalar field models include
Quintessence[4-14], K-essence[15], Tachyon[16], Phantom[17-20],
Quintom[21] and so on.
\par On the other hand, the role of tachyon field field in string theory in cosmology has
been widely studied[22]. It shows that the tachyon can be described
by a B-I type Lagrangian resulting from sting theory. Nonlinear
Born-Infeld scalar field theory,
 firstly proposed by W.Heisenberg in order to describe the process
 of meson multiple production connected with strong field
 regime[23], is discussed in cosmology recently[24]. It shows that
 the lagrangian density of this B-I scalar field posses some
 interesting characteristics[25,26]. In Ref[25], the author showed that a
 horizon exists for a large class of solution in which the scalar
 field is finite though this horizon is singular. Naked
 singularities with everywhere well-behaved scalar field in
 another class of solution have also been found. Lately the
 quantum cosmology with the B-I scalar field has been
 considered[27], in the extreme limits of small and large
 cosmological scale factor the wave function of the universe was
 found by applying the methods developed by Vilenkin, Hartle and
 Hawking. The result has suggested a non-zero positive
 cosmological constant with largest probability, which is
 consistent with current observational data. The classical
 wormhole solution and wormhole wavefunction with the B-I scalar
 filed has been obtained in Ref[28]. The phantom
 cosmology based on B-I scalar field with a special potential has been considered in Ref[29].
 The result shows that the universe will evolve to a de-sitter
 like attractor regime in the future and the phantom B-I scalar
 field can survive till today without interfering with the
 nucleosynthesis of the standard model.
\par Recently, many authors have investigated the evolution
of dark energy models in the $\omega-\omega'$ plane[30,31], where
$\omega'$ is the time variation of $\omega$ with respect to $N$.
According to different regions in the $\omega-\omega'$ phase plane,
these models can be classified two types which are call "thawing"
and "freezing" models. Based on this consideration, we study the
cosmological dynamics of B-I type phantom model with the potential
$u(\varphi)=A\varphi e^{-\beta\varphi}$ in the $\omega-\omega'$
plane and investigate the properties of attractor solutions.
According to the obtained B-I field equation of motion, we analyze
the critical points and find the critical point with $\omega\sim-1$
is the late-time attractor, where B-I field becomes ultimately
frozen, as shown in the Fig.1. \vskip 0.2 in
\begin{flushleft}\textbf{2. B-I Type Phantom Model}\end{flushleft}
In 1934[32], Born and Infeld put forward a theory of non-linear
electromagnetic field. The lagrangian density is
\begin{equation}\displaystyle
L_{BI}=b^2\left[1-\sqrt{1-(\frac{1}{2b^2})F_{\mu\nu}F^{\mu\nu}}~\right]\end{equation}
The lagrangian density for a B-I type scalar field is
\begin{equation}\displaystyle L_S=\frac{1}{\eta}\left[1-\sqrt{1-\eta g^{\mu\nu}\varphi_{,~\mu}\varphi_{,~\nu}}~\right]\end{equation}
Eq.(2) is equivalent to the tochyon lagrangian
$[-V(\varphi)\sqrt{1-g^{\mu\nu}\varphi_{,~\mu}\varphi_{,~\nu}}+\Lambda]$
if $\displaystyle V(\varphi)=\frac{1}{\eta}$ and cosmological
constant $\displaystyle \Lambda=\frac{1}{\eta}$ ($\displaystyle
\frac{1}{\eta}$ is two times as "critical" kinetic energy of
$\varphi$ field). The lagrangian (2) possesses some interesting
characteristics, it is exceptional in the sense that shock waves do
not develop under smooth or continuous initial conditions and
because nonsingular scalar field solution can be generated[33]. When
$\eta \rightarrow 0$, by Taylor expansion, Eq.(2) approximates to
the lagrangian of linear scalar field.
\begin{equation}\lim_{\eta\rightarrow 0}L_S=\frac{1}{2}g^{\mu\nu}\varphi_{,~\mu}\varphi_{,~\nu}\end{equation}
Now we consider the Lagrangian of phantom model with a potential
$u(\varphi)$ in spatially homogeneous scalar field, Eq.(2) becomes
\begin{equation}\end{equation}$$L_{ph}=\frac{1}{\eta}\left[1-\sqrt{1+\eta \dot{\varphi}^2}~\right]-u(\varphi)$$
In the spatially flat Robertson-Walker metric
$ds^2=dt^2-a^2(t)(dx^2+d^2y+d^2z)$, Einstein equation
$G_{\mu\nu}=KT_{\mu\nu}$ in phantom dominated epoch, can be written
as
\begin{equation}H^2=\frac{1}{3}\rho_\varphi\end{equation}
\begin{equation}\dot{\rho}_\varphi+3H(\rho_\varphi+p_\varphi)=0\end{equation}
where $\rho_\varphi$ and $p_\varphi$ are the effective energy
density and effective pressure respectively, and we work in units
$8\pi G=1$. The Energy-moment tensor is
\begin{equation}\displaystyle T^\mu_\nu=-\frac{g^{\mu\rho}\varphi_{,~\nu}\varphi_{,~\rho}}{\sqrt{1+\eta g^{\mu\nu}\varphi_{,~\mu}\varphi_{,~\nu}}}-\delta^\mu_\nu L_{ph}\end{equation}
From Eq.(7), we have
\begin{equation}\rho_\varphi=T^0_0=\frac{1}{\eta\sqrt{1+\eta \dot{\varphi}^2}}-\frac{1}{\eta}+u\end{equation}
\begin{equation}p_\varphi\equiv\omega_\varphi\rho_\varphi=-T^1_1=-T^2_2=-T^3_3=\frac{1}{\eta}-\frac{\sqrt{1+\eta \dot{\varphi}^2}}{\eta}-u\end{equation}
From Eqs.(8)(9), we get
\begin{equation}\rho_\varphi+p_\varphi=-\frac{\dot{\varphi}^2}{\sqrt{1+\eta \dot{\varphi}^2}}\end{equation}
It is clear that the equation of static $\omega<-1$ is completely
confirmed by Eq.(10)and it accords with the recent analysis of
observation data. \vskip 0.2 in  \begin{flushleft}\textbf{3.
Dynamics of B-I Type Phantom Model In $\omega-\omega'$
Plane}\end{flushleft}
\par  We make such a definition $N=lna$ which results in the no-dimension of scale factor $a$.  From Eqs.(5)(6)(9) and (10), we have
\begin{equation}\frac{d\rho_\varphi}{dN}=-3(\omega_\varphi+1)\rho_\varphi\end{equation}
\begin{equation}\frac{d\varphi}{dN}=\frac{\dot{\varphi}}{\sqrt{\frac{\rho_\varphi}{3}}}\end{equation}
\begin{equation}(\omega_\varphi+1)\rho_\varphi=-\frac{\dot{\varphi}^2}{\sqrt{1+\eta\dot{\varphi}^2}}\end{equation}
Now, we defining a function
\begin{equation}\Delta(a)\equiv \frac{d[ln~u(\varphi)]}{d(ln\rho_\varphi)}=1+\frac{d(ln(1-\omega_\varphi))}{d(ln\rho_\varphi)}=1+\frac{d\omega_\varphi/dN}{3(1-\omega^2_\varphi)}\end{equation}
So, we can rewrite Eq.(14) as
\begin{equation}\frac{d\omega_\varphi}{dN}=[3(1-\omega^2_\varphi)]\times[\Delta-1]\end{equation}
where
\begin{equation}\Delta=\frac{\pm u'/u}{\frac{1}{\rho_\varphi}\cdot\frac{d\rho_\varphi/dN}{d\varphi/dN}}\end{equation}
and the sign $"'"$ denotes the derivative of $u(\varphi)$ with
respect to $\varphi$. Eq.(15) is the scalar field equation of motion
in B-I type phantom model. For the B-I field rolling down its
potential, the $\pm$ sign before $u'(\varphi)$ corresponds to
$u'(\varphi)>0$ or $u'(\varphi)<0$. When $\eta\rightarrow0$, the
scalar field equation of motion in B-I type phantom model will tend
to be the canonical case.
\par Now we consider the potential $u(\varphi)=A\varphi e^{-\beta\varphi}$. The system
decided by Eq.(15) admits these critical points $\omega_\varphi=1$,
$\omega_\varphi\sim-1$, $\Delta=1$. When $\Delta=1$, $\omega_\sigma$
varies very slowly where the B-I field is tracking and the ratio of
kinetic energy to potential energy of the B-I field becomes a
constant. When $\omega_\varphi\sim-1$, it corresponds to a late time
attractor where the B-I field becomes ultimately frozen, as shown in
the Fig.1. When we set different initial conditions, the critical
point always tends $\omega_\sigma\sim-1$, as shown in Fig.2. Fig.3
shows the evolution of the state equation of state $\omega_\varphi$
oscillates at the beginning and tends to be -1 with respect to $N$
eventually.

\vskip 0.15 in
\begin{center}
\begin{minipage}{0.5\textwidth}
\includegraphics[scale=0.8]{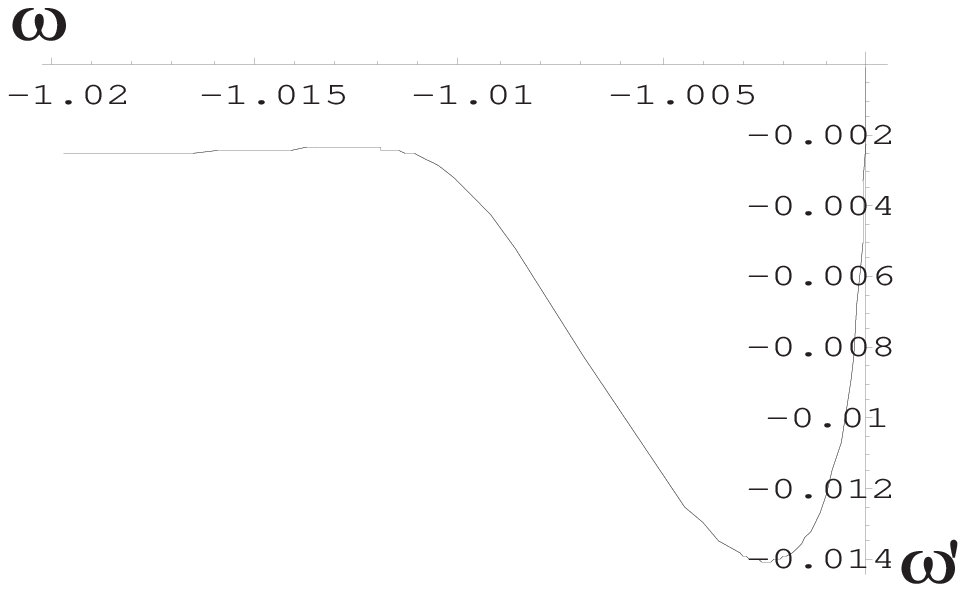}
{\small Fig.1 The attractive behavior of B-I model with potential
$u(\varphi)=A\varphi e^{-\beta\varphi}$ in the $\omega-\omega'$
plane. We set $\eta=10^{-4}$, $\beta=0.5$ and $A=1.0$.}
\end{minipage}
\end{center}

\vskip 0.3 in
\begin{center}
\begin{minipage}{0.5\textwidth}
\includegraphics[scale=0.8]{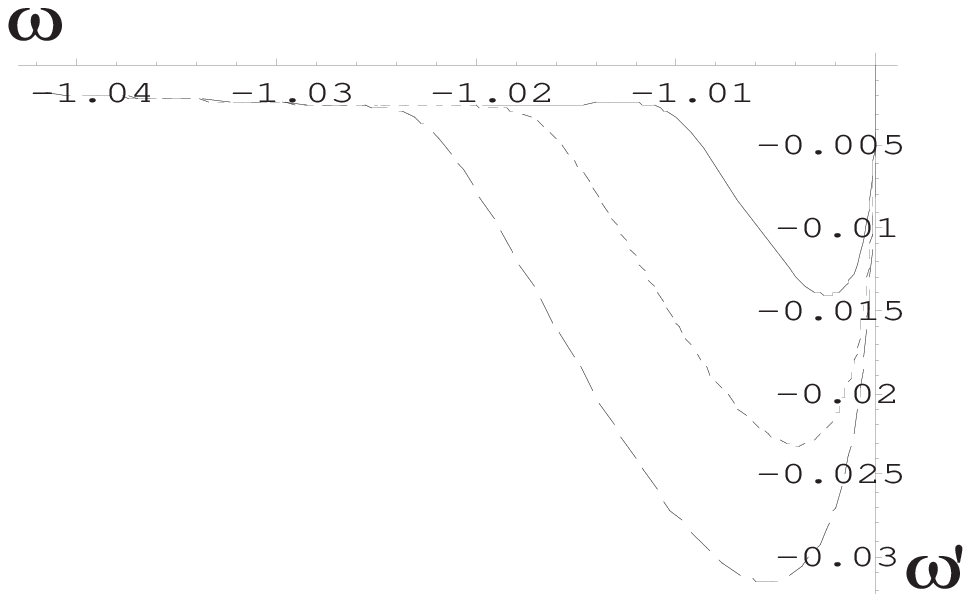}
{\small Fig.2 The comparison of attractive behavior of B-I model
with potential $u(\varphi)=A\varphi e^{-\beta\varphi}$ for different
initial conditions in the $\omega-\omega'$ plane.}
\end{minipage}
\end{center}

\vskip 0.3 in
\begin{center}
\begin{minipage}{0.5\textwidth}
\includegraphics[scale=0.8]{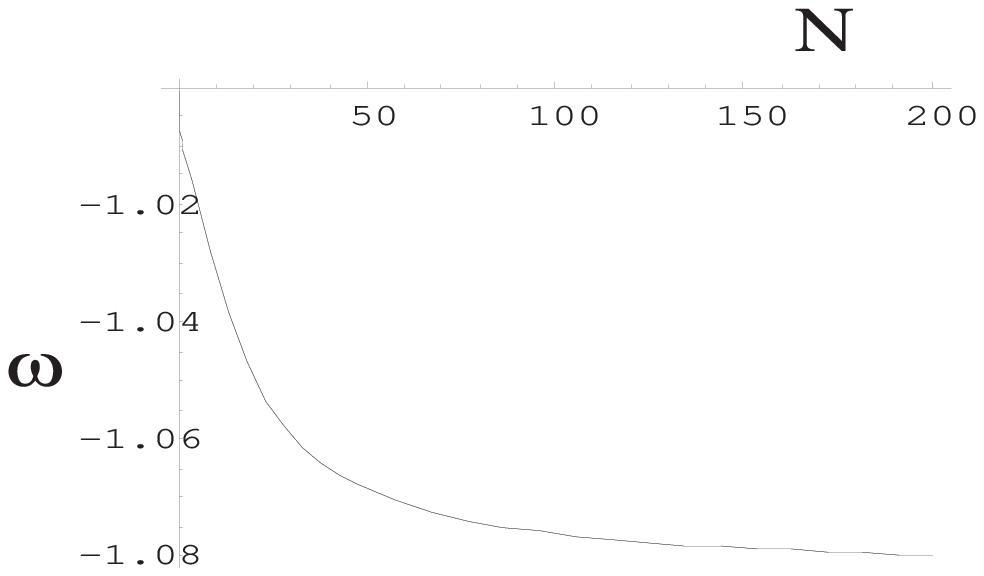}
{\small Fig.3 The evolutive behavior of equation of state parameter
$\omega_\varphi$. $\omega_\varphi-1$ which corresponds to a late
time attractor, where the B-I phantom field becomes ultimately
frozen. $\mid1+\omega_\varphi\mid=\mid1+(-1.08)\mid=0.08> 0.02$},
which satisfies the limit of "freezing" model.
\end{minipage}
\end{center}

\vskip 0.3 in
\begin{flushleft}\textbf{{4. Conclusions}}\end{flushleft}
\par Caldwell and Linder[29] classified the quintessence models into two
types "thawing" and "freezing" model according to the different
regions in $\omega-\omega'$ plane and gave the limit of
quintessence[28]: $1+\omega\geq0.004$ for "thawing" model and
$1+\omega\geq0.01$. We consider the dynamics of B-I phantom model in
the $\omega-\omega'$ plane for the first time. In this paper, we
investigate the cosmological dynamics of B-I phantom model with
potential $A\varphi e^{-\beta\varphi}$ in the $\omega-\omega'$ plane
and examine the evolution of equation of state parameter with
respect to $N$. These numerical results show that the critical point
with $\omega\sim-1$ is a late-time attractor, where B-I field
becomes ultimately frozen and $\omega_\varphi\sim-1$. The existence
of the attractor solution prevents the phantom energy from
increasing up to infinite in a finite time, therefor the presence of
phantom energy dose not lead to a cosmic doomsday in a theory with
late time attractor. We also consider the effect of the initial
condition on the property of attractor solution.
$\mid1+\omega_\varphi\mid=\mid1+(-1.08)\mid=0.08> 0.02$, which
satisfies the limit of "freezing" model. So, we know our B-I model
belongs to "freezing" model, where the B-I field which was already
rolling up towards its potential maximum $\frac{A}{\beta e}$ which
is decided by $u'(\varphi)=0$, prior to the onset of acceleration,
but which slows up and creeps to a halt as it comes to dominate the
Universe. \vskip 0.3 in
\begin{flushleft}\textbf{Acknowledgements}\end{flushleft}
This work is partially supported by National Nature Science
Foundation of China under Grant No.10573012 and Shanghai Municipal
Science and Technology Commission No.04dz05905.

\begin{flushleft}{\noindent\bf References}
 \small{
\item{1.}{ A. G. Riess et al., \textit{Astrophys. J}\textbf{607}, 665(2004);
\\\hspace{0.15 in}A. G. Riess, \textit{Astron. J}\textbf{116}, 1009(1998);
\\\hspace{0.15 in}S. Perlmutter et al., \textit{Astrophys. J}\textbf{517}, 565(1999);
\\\hspace{0.15 in}N. A Bahcall et al., \textit{Science}\textbf{284}, 1481(1999).}
\item{2.}{ D. N. Spergel et al., Astrophys. J. Suppl\textbf{148}, 175(2003).}
\item{3.}{ P. de Bernardis et al., arXiv:astro-ph/0105296;
\\\hspace{0.15 in}R. Stompor et al., arXiv:astro-ph/0105062;
\\\hspace{0.15 in}C. Pryke et al., arXiv:astro-ph/0104490.}
\item {4.}{ J. S. Bagla, H. K. Jassal and T. Padmamabhan, \textit{Phys. Rev. D}\textbf{67}, 063504(2003).}
\item {5.}{ L. Amendola, M. Quartin, S. Tsujikawa and I. Waga, \textit{Phys.Rev.D}\textbf{74}, 023525(2006);
\\\hspace{0.15 in}S. Nojiri, S. D. Odintsov and M. Sasaki, \textit{Phys.Rev. D}\textbf{70} 043539(2004);
\\\hspace{0.15 in}S. Nojiri and S. D. Odintsov, arXiv:hep-th/0601213;
\\\hspace{0.15 in}S. Nojiri and S. D. Odintsov \textit{Phys. Lett. B}\textbf{639}, 144(2006);
\\\hspace{0.15 in}S. Nojiri, S. D. Odintsov and H. Stefancic[arXiv:hep-th/0608168];
\\\hspace{0.15 in}S. Nojiri and S. D. Odintsov, \textit{Phys. Lett. B}\textbf{562}, 147(2003)[arXiv:hep-th/0303117];
\\\hspace{0.15 in}S. Nojiri and S.D. Odintsov, \textit{Phys. Rev. D}\textbf{70}, 103522(2004)[arXiv:hep-th 0408170];
\\\hspace{0.15 in}S. Nojiri, S. D. Odintsov and S. Tsujikawa, arXiv:hep-th/0501025;
\\\hspace{0.15 in}S. Nojiri and S. D. Odintsov, \textit{Phys. Rev. D}\textbf{7}2, 023003(2005)[arXiv:hep-th/0505215];
\\\hspace{0.15 in}S. Nojiri and S. D. Odintsov, arXiv:hep-th 0506212;
\\\hspace{0.15 in}S. Nojiri and S. D. Odintsov, arXiv:hep-th/0611071;
\\\hspace{0.15 in}E. Elizalde, S. Nojiri and S. D. Odintsov, arXiv:hep-th/0405034;
\\\hspace{0.15 in}B. Boisseau et al., \textit{Phys. Rev. Lett}\textbf{85}, 2236(2000)[arXiv:gr-qc/0001066];
\\\hspace{0.15 in}G. Esposito-Farese and D. Polarski, \textit{Phys. Rev. D}\textbf{63}, 063504(2001)[arXiv:gr-qc/0009034];
\\\hspace{0.15 in}Xin Zhang, \textit{Mod. Phys. Lett. A}\textbf{20}, 2575(2005)[arXiv:astro-ph/0503072];
\\\hspace{0.15 in}Xin Zhang, \textit{Phys. Lett. B}\textbf{611}, 1(2005)[arXiv:astro-ph/0503075];
\\\hspace{0.15 in}M. R. Setare, \textit{Phys. Lett. B}\textbf{642},1(2006)[arXiv:hep-th/0609069];
\\\hspace{0.15 in}M. R  Setare, arXiv:hep-th/0609104;
\\\hspace{0.15 in}M. R. Setare, arXiv:hep-th/0610190.}
\item{6.}{ C. Wetterich \textit{Nucl. Phys. B}\textbf{302}, 668(1998);
\\\hspace{0.15 in}E. J. Copeland, M. Sami and S. Tsujikawa, arXiv:hep-th/0603057;
\\\hspace{0.15 in}P. G. Ferreira and M. Joyce \textit{Phys. Rev. D}\textbf{58}, 023503(1998);
\\\hspace{0.15 in}J. Frieman, C. T. Hill, A. Stebbinsand and I.Waga, \textit{Phys. Rev. Lett}\textbf{75}, 2077(1995);
\\\hspace{0.15 in}P. Brax and J. Martin, \textit{Phys. Rev. D}\textbf{61}, 103502(2000);
\\\hspace{0.15 in}T. Barreiro, E. J. Copeland and N. J. Nunes, \textit{Phys. Rev. D}\textbf{61}, 127301(2000);
\\\hspace{0.15 in}I. Zlatev, L. Wang and P. J. Steinhardt \textit{Phys. Rev. Lett} \textbf{82}, 896(1999);
\\\hspace{0.15 in}S. Nojiri and S. D. Odintsov, arXiv:hep-th/0306212;
\\\hspace{0.15 in}S. Nojiri and S. D. Odintsov, arXiv:hep-th/0601213.}
\item{7.}{T. Padmanabhan, and T. R. Choudhury, \textit{Phys. Rev. D}\textbf{66}, 081301(2002).}
\item{8.}{A. Sen, \textit{JHEP} \textbf{0204}, 048(2002).}
\item{9.}{C. Armendariz-Picon, T. Damour and V. Mukhanov, \textit{Phys. Lett. B}\textbf{458}, 209(1999).}
\item{10.}{A. Feinstein, \textit{Phys. Rev. D}\textbf{66}, 063511(2002);
\\\hspace{0.17 in}M. Fairbairn and M. H. Tytgat, \textit{Phys. Lett. B}\textbf{546} 1(2002).}
\item{11.}{A. Frolov, L. Kofman and A. Starobinsky, \textit{Phys.Lett.B} \textbf{545}, 8(2002);
\\\hspace{0.17 in}L. Kofman and A. Linde, \textit{JHEP} \textbf{0207}, 004(2002).}
\item{12.}{C. Acatrinei and C. Sochichiu, \textit{Mod. Phys. Lett. A}\textbf{18}, 31(2003);
\\\hspace{0.17 in}S. H. Alexander, \textit{Phys. Rev. D}\textbf{65}, 0203507(2002).}
\item{13.}{T. Padmanabhan, \textit{Phys. Rev. D}\textbf{66}, 021301(2002).}
\item{14.}{A. Mazumadar, S. Panda and A. Perez-Lorenzana, \textit{Nucl. Phys. B}\textbf{614}, 101(2001);
\\\hspace{0.17 in}S. Sarangi and S. H. Tye, \textit{Phys. Lett. B}\textbf{536}, 185(2002).}
\item{15.}{C. Armend\'{a}riz-Pic\'{o}n, V. Mukhanov and P. J. Steinhardt, \textit{Phys.Rev.Lett}\textbf{85}, 4438(2000);
\\\hspace{0.17 in}C. Armend\'{a}riz-Pic\'{o}n, V. Mukhanov and P. J. Steinhardt, \textit{Phys. Rev. D}\textbf{63}, 103510(2001);
\\\hspace{0.17 in}T. Chiba, \textit{Phys.Rev.D}\textbf{66}, 063514(2002);
\\\hspace{0.17 in}T. Chiba, T. Okabe and M. Yamaguchi, \textit{Phys. Rev. D}\textbf{62}, 023511(2000);
\\\hspace{0.17 in}M. Malquarti, E. J. Copeland, A. R. Liddle and M. Trodden, \textit{Phys. Rev. D}\textbf{67}, 123503(2003);
\\\hspace{0.17 in}R. J. Sherrer, \textit{Phys. Rev. Lett}\textbf{93)}, 011301(2004);
\\\hspace{0.17 in}L. P. Chimento, \textit{Phys. Rev. D}\textbf{69}, 123517(2004);
\\\hspace{0.17 in}A. Melchiorri, L. Mersini, C. J. Odman and M. Trodden, \textit{Phys. Rev. D}\textbf{68}, 043509(2003).}
\item{16.}{A. Sen, \textit{JHEP} \textbf{0207}, 065(2002);
\\\hspace{0.17 in}M. R. Garousi, \textit{Nucl. Phys. B}\textbf{584}, 284(2000);
\\\hspace{0.17 in}M. R. Garousi, \textit{JHEP} \textbf{0305}, 058(2003);
\\\hspace{0.17 in}E. A. Bergshoeff, M. de Roo, T. C. de Wit, E. Eyras and S. Panda,\textit{ JHEP} \textbf{0005}, 009(2000);
\\\hspace{0.17 in}J. Kluson, \textit{Phys. Rev. D}\textbf{62}, 126003(2000);
\\\hspace{0.17 in}G. W. Gibbons, \textit{Phys. Lett. B}\textbf{537}, 1(2002);
\\\hspace{0.17 in}M. Sami, P. Chingangbam and T. Qureshi, \textit{Phys. Rev. D}\textbf{66}, 043530(2002);
\\\hspace{0.17 in}M. Sami, \textit{Mod. Phys. Lett. A}\textbf{18}, 691(2003);
\\\hspace{0.17 in}Y. S. Piao, R. G. Cai, X. m. Zhang and Y. Z. Zhang, \textit{Phys. Rev. D}\textbf{66},121301(2002);
\\\hspace{0.17 in}E. J Copeland et al., \textit{Phys. Rev. D}\textbf{71}, 043003(2005)[arXiv:hep-th/0411192].}
\item{17.}{Z. G. Huang, H. Q. Lu and W. Fang, \textit{Class. Quant. Grav}\textbf{23}, 6215(2006)[arXiv:hep-th/0604160];
\\\hspace{0.17 in}Z. G. Huang and H. Q. Lu, \textit{Int. J. Mod. Phys. D}\textbf{15}, 1501(2006);
\\\hspace{0.17 in}H. Q. Lu, \textit{Int. J. Mod. Phys. D}\textbf{14}, 355(2005)[arXiv:hep-th/0312082];
\\\hspace{0.17 in}H. Q. Lu, Z. G. Huang, W. Fang and P. Y. Ji, arXiv:hep-th/0504038;
\\\hspace{0.17 in}W. Fang, H. Q. Lu and Z. G. Huang, arXiv:hep-th/0606032.}
\item{18.}{X. Z. Li and J. G. Hao, \textit{Phys. Rev. D}\textbf{69}, 107303(2004).}
\item{19.}{T. Chiba, T. Okabe and M. Yamaguchi, \textit{Phys. Rev. D}\textbf{62}, 023511(2002).}
\item{20.}{P. Singh, M. Sami and N. Dadhich, \textit{Phys.Rev. D}\textbf{68}, 023522(2003);
\\\hspace{0.17 in}J. Kujat, R. J. Scherrer and A. A. Sen, \textit{Phys. Rev. D}\textbf{74}, 083501(2006)[arXiv:astro-ph/0606735].}
\item{21.}{W. Hao, R. G. Cai and D. F. Zeng, \textit{Class.Quant.Grav}\textbf{22}, 3189(2005);
\\\hspace{0.17 in}Z. K. Guo, Y. S. Piao, X. M. Zhang, Y.Z. Zhang, \textit{Phys.Lett. B}\textbf{608}, 177(2005);
\\\hspace{0.17 in}B. Feng, arXiv:astro-ph/0602156.}
\item{22.}{S. M. Carroll, M. Hoffman and M. Trodden, \textit{Phys. Rev. D}\textbf{68}, 023509(2003);
\\\hspace{0.17 in}J. G. Hao and X. Z. Li, \textit{Phys. Rev. D}\textbf{68}, 043501(2003);
\\\hspace{0.17 in}S. Mukohyama, \textit{Phys. Rev. D}\textbf{66}, 024009(2002);
\\\hspace{0.17 in}T. Padmanabhan, \textit{Phys. Rev. D}\textbf{66}, 021301(2002);
\\\hspace{0.17 in}M. Sami and T. Padamanabhan, \textit{Phys. Rev. D}\textbf{67}, 083509(2003);
\\\hspace{0.17 in}G. Shiu and I. Wasserman, \textit{Phys. Lett. B}\textbf{541}, 6(2002);
\\\hspace{0.17 in}T. Brunier, V. K. Onemli, and R. P. Woodard, \textit{Class. Quant. Grav}\textbf{2}2,59(2005)[arXiv:gr-qc/0408080];
\\\hspace{0.17 in}V. K. Onemli and R. P. Woodard, \textit{Phys. Rev. D}\textbf{70}, 107301(2004)[arXiv:gr-qc/0406098];
\\\hspace{0.17 in}V. K. Onemli and R. P. Woodard, \textit{Class. Quant. Grav}\textbf{19}, 4607(2002)[arXiv:gr-qc/0204065].}
\item {23.}{W. Heisenberg, \textit{Z. Phys.}\textbf{133}, 79(1952);
\\\hspace{0.17 in}W. Heisenberg, \textit{Z. Phys.}\textbf{126}, 519(1949);
\\\hspace{0.17 in}W. Heisenberg, \textit{Z. Phys.}\textbf{113},61(1939).}
\item{24.}{G. W. Gibbons and C. A. R. Herdeiro, \textit{Phys. Rev. D}\textbf{63}, 064006(2001);
\\\hspace{0.17 in}G. W. Gibbons \textit{Rev. Mex. Fis.}\textbf{49}S1, 19(2003);
\\\hspace{0.17 in}V. V. Dyadichev, D. V. Gal'tsov and A. G. Zorin, \textit{Phys. Rev. D}\textbf{65}, 084007(2002);
\\\hspace{0.17 in}D. N. Vollick, \textit{Gen. Rel. Grav}.\textbf{35}, 1511(2003).}
\item{25.}{H. P. de Oliveira, \textit{J. Math. Phys}.\textbf{36}, 2988(1995).}
\item{26.}{T. Taniuti, \textit{Prog. Theor. Phys.}(kyoto) Suppl\textbf{9}, 69(1958).}
\item{27.}{H. Q. Lu, T. Harko and K. S. Cheng, \textit{Int. J. Modern. Phys. D}\textbf{8}, 625(1999);
\\\hspace{0.17 in}Z. G. Huang, H. Q. Lu and P. P. Pan, \textit{Astrophysics and Space Science}\textbf{295}, 493(2005).}
\item{28.}{H. Q. Lu et al., \textit{Int. J. Theor. Phys}\textbf{42}, 837(2003).}
\item{29.}{W. Fang, H. Q. Lu, Z. G. Huang and K. F. Zhang, \textit{Int. J. Mod. Phys. D}\textbf{15}, 199(2006)[arXiv:hep-th/0409080].}
\item{30.}{R. R. Caldwell and E. V. Linder, \textit{Phys. Rev. Lett.}\textbf{95}, 141301(2005).}
\item{31.}{S. A. Bludman and M. Roos, \textit{Phys.Rev. D}\textbf{65}, 043503(2002);
\\\hspace{0.17 in}Z. K. Guo, Y. S. Piao, X. M. Zhang and Y. Z. Zhang, arXiv:astro-ph/0608165.}
\item{32.}{M. Born and Z. Infeld, \textit{Proc. Roy. Soc. A}\textbf{144}, 425(1934).}
\item{33.}{H. P.de Oliveira, \textit{J. Math. Phys.}\textbf{36}, 2988(1995).}
}
\end{flushleft}
\end{document}